\documentclass[authoryear,a4paper,review]{elsarticle}
\journal{arXiv}
\pdfoutput=1

\usepackage{natbib,graphicx,amsmath,amssymb,ifthen,hyperref,setspace,soul,xfrac}
\usepackage[left=1in,right=1in,top=1.4in,bottom=1.4in]{geometry}

\hypersetup{
	pdfauthor={Jaehyuk Choi, Byoung Ki Seo},
	pdfkeywords={stochastic volatility, normal SABR model, numerical quadrature},
	colorlinks=true,
	linkcolor=red,
	citecolor=blue,
	urlcolor=blue,
	bookmarksnumbered=true,
}

\newcommand{\rhoc}{\rho_\ast}

\newcommand{\vov}{\nu}
\newcommand{\vovn}{\xi}
\newcommand{\vovnsq}{\xi^2}
\newcommand{\ulam}{u_\lambda}
\newcommand{\distequal}{\,{\buildrel d \over =}\,}

\newcommand{\sgn}{\mathrm{sgn}}
\newcommand{\norm}{\textsc{n}}

\newcommand{\drift}{\lambda}
\newcommand{\half}{\sfrac{1}{2}}
\newcommand{\Zdrift}[1][-\half]{\ifthenelse{\equal{#1}{0}}{Z}{Z^{[#1]}}}
\newcommand{\ExpBM}[1][-\half]{\ifthenelse{\equal{#1}{0}}{A}{A^{[#1]}}}

\newcommand{\fwd}{F_0}

\newcommand{\qtext}[2][\quad]{#1\text{#2}#1}

\newproof{pf}{Proof}

\begin{document}
\begin{frontmatter}
\title{Option pricing under the normal SABR model\\ with Gaussian quadratures}

\author[phbs]{Jaehyuk Choi\corref{corrauthor}}
\ead{jaehyuk@phbs.pku.edu.cn}

\author[unist]{Byoung Ki Seo}
\ead{bkseo@unist.ac.kr}

\cortext[corrauthor]{Corresponding author \textit{Tel:} +86-755-2603-0568; \textit{Address:} Rm 755, Peking University HSBC Business School, University Town, Nanshan, Shenzhen 518055, China}

\address[phbs]{Peking University HSBC Business School,\\ University Town, Nanshan, Shenzhen 518055, China}
\address[unist]{School of Business Administration, Ulsan National Institute of Science and Technology}

\begin{abstract}
The stochastic-alpha-beta-rho (SABR) model has been widely adopted in options trading. In particular, the normal ($\beta=0$) SABR model is a popular model choice for interest rates because it allows negative asset values. The option price and delta under the SABR model are typically obtained via asymptotic implied volatility approximation, but these are often inaccurate and arbitrage-able. Using a recently discovered price transition law, we propose a Gaussian quadrature integration scheme for price options under the normal SABR model. The compound Gaussian quadrature sum over only 49 points can calculate a very accurate price and delta that are arbitrage-free.
\end{abstract}
\date{November 16, 2022}

\end{frontmatter}

\section{Introduction}
\noindent
The stochastic-alpha-beta-rho (SABR) model~\citep{hagan2002sabr} has been widely adopted in the financial industry for the pricing and risk management of European options, owing to its intuitive and parsimonious parameterization. It has been a standard practice for practitioners to obtain the option price and delta from the asymptotic approximation of implied volatility~\citep{hagan2002sabr}, but the approximation loses accuracy and allows arbitrage as the variance of volatility becomes large. Despite numerous attempts to improve implied volatility approximation~\citep{obloj2007fine,paulot2015asym,lorig2017lsv,yang2017cev-sabr,choi2021sabrcev}, it does not seem possible to obtain an approximation accurate for all parameter ranges. To date, there are several \textit{full-scale} methods for pricing the SABR model: Monte--Carlo simulations~\citep{chen2012low,leitao2017one,leitao2017multi,cai2017sabr,choi2019nsvh,cui2021efficient}, finite difference methods~\citep{park2014sabr,vonsydow2019benchop}, and continuous-time Markov chains~\citep{cui2018ctmc-sabr}. These approaches require heavy computation and complex implementation compared with the approximation approach. Therefore, they are difficult to implement in practice.

We contribute to the literature by proposing a novel and efficient pricing scheme under the normal ($\beta=0$) SABR model. Despite being a special case, the normal SABR model has gained attention for modeling interest rates owing to its flexibility to allow negative asset prices. The normal model based on the arithmetic Brownian motion (BM) has long been used in fixed income markets, as opposed to the Black--Scholes model based on geometric BM.\footnote{See \citet{choi2022bachelier} for a survey of the normal (Bachelier) model.} The normal SABR model is a natural extension of the normal model with stochastic volatility that exhibits a volatility smile. \citet{antonov2015mixing} adopt the normal SABR model as a key component of the mixture approach for modeling a low-interest-rate environment. \citet{choi2019nsvh} propose the hyperbolic normal stochastic volatility (NSVh) model as a broader class of normal stochastic volatility models.

Based on the price transition law of \citet{choi2019nsvh}, we express the option price as a double sum of compound Gaussian quadratures consisting of Gauss--Hermite and Gauss--Laguerre quadratures. Our scheme is highly efficient in the sense that quadrature points with only 49 (i.e., $7\times 7$) nodes can produce very accurate option values. Unlike with the asymptotic implied volatility approach, the resulting prices obtained with our approach are arbitrage-free. Using a similar algorithm, we can also evaluate the option delta (equivalently, the probability distribution). Our study extends the previous literature on the normal SABR model. It compliments \citet{choi2019nsvh} by showing that the price transition law is useful for deterministic pricing as well as Monte--Carlo simulation. It extends the integral representations of the normal SABR model previously observed by \citet{korn2013exact} and \citet{antonov2015mixing}, as we provide an alternative representation and a practical numerical scheme together. The study also extends the stochastic volatility benchmark proposed by \citet{vonsydow2019benchop}. Given the availability of highly accurate option values, our method can serve as a testing benchmark for general SABR pricing methods proposed in the future. 

The remainder of this paper is organized as follows. Section~\ref{s:model} reviews the normal SABR model. Section~\ref{s:quad} introduces the proposed quadrature scheme. Section~\ref{s:num} presents the numerical results. Finally, Section~\ref{s:conc} concludes the paper.

\section{Normal SABR model} \label{s:model}
\noindent
The SABR model~\citep{hagan2002sabr} is a stochastic volatility model specified by
\begin{equation}
\frac{dF_t}{F_t^\beta} = \sigma_t \, \rho dW_t \quad \text{and}\quad \frac{d\sigma_t}{\sigma_t} = \vov\, dZ_t,
\end{equation}
where $F_t$ and $\sigma_t$ are the processes for the forward price and volatility, respectively; $\vov$ is the volatility of volatility; $\beta\in[0,1]$ is the elasticity parameter; and
$W_t$ and $Z_t$ are the BMs correlated by $dW_t dZ_t = \rho\, dt$.
To simplify the notations for the remainder of this paper, we also denote 
\begin{equation} \label{eq:rhoc}
\rhoc = \sqrt{1-\rho^2} \qtext{and} \vovn = \frac12 \vov\sqrt{T},
\end{equation}
where $T$ is the option expiry.

The normal SABR model is the case with $\beta=0$. Unlike when $\beta>0$, the asset price $F_t$ can freely become negative in the normal SABR model, requiring no boundary condition at the origin.\footnote{In some studies~\citep{antonov2015mixing}, the model is explicitly referred to as the normal \textit{free-boundary} SABR. When $0<\beta< 1$, the absorbing boundary condition is imposed at the origin, resulting in a mass at zero. See \citet{gulisashvili2018mass,chen2019feeling,choi2021note} for details.} 
Here, we are concerned with the undiscounted price and delta of the call option with strike price $K$ and expiry $T$:
$$ C(K) = E([F_T - K]^+) \qtext{and} D(K) = \frac{\partial C(K)}{\partial F_0}.$$
Since $F_0$ and $K$ affect the option value only through moneyness, $F_0-K$, under the normal SABR model, the option delta is also equal to the complementary probability distribution of $F_T$:
$$ 
D(K) = \frac{\partial C(K)}{\partial F_0} = -\frac{\partial C(K)}{\partial K} = \text{Prob}(F_T\ge K).
$$

\subsection{\citet{hagan2002sabr}'s normal volatility approximation} \label{ss:volapprox}
\noindent
\citet{hagan2002sabr} present the celebrated implied Black--Scholes volatility of the SABR model, from which the option price can be quickly computed using the Black--Scholes formula. In their study, \citet{hagan2002sabr} first derive the implied normal volatility, even for all $\beta\in[0,1]$, because they use perturbation from normal diffusion. They then convert the derived volatility to the equivalent Black--Scholes volatility using another approximation. However, for the normal SABR model, normal volatility is more relevant, as has been discussed. Therefore, the normal volatility approximation in the intermediate step~\citep[(A.59)]{hagan2002sabr} is preferred among practitioners, as it avoids further approximation error. The implied normal volatility for $\beta=0$ is given by
\begin{equation} \label{eq:hagan}
\sigma_\norm(K) = \sigma_0 \frac{z}{x(z)} \left(1+\frac{2-3\rho^2}{6} \vovnsq\right)\\ \qtext{for} z = \frac{\vov}{\sigma_0}(K - F_0)
\end{equation}
where $K$ is the strike price, $\vovn$ is given in Eq.~\eqref{eq:rhoc}, and $x(z)$ is defined by\footnote{At and near $z=0$, $z/x(z)$ should be evaluated as 1 from Taylor's expansion near $z=0$:
$$
\frac{x(z)}{z} = 1 - \frac{\rho}{2}\,z + \frac{3\rho^2-1}{6}\,z^2 - \frac{(5\rho^2 - 3)\rho}8\, z^3
+ \;\cdots.
$$}

\begin{equation} \label{eq:Hxv}
x(z) = \log\left( \frac{V(z)+z+\rho}{1+\rho} \right) \qtext{for} V(z) =\sqrt{1+2\rho z+z^2}.
\end{equation}
The option price is obtained by plugging $\sigma_\norm(K)$ into the Bachelier option formula:
\begin{equation} \label{eq:p_hagan}
C(K) \approx (F_0 - K) N(d_\norm) + \sigma_\norm\sqrt{T} \, n(d_\norm)
\qtext{for} d_\norm = \frac{F_0 - K}{\sigma_\norm(K) \sqrt{T}},
\end{equation}
where $n(z)$ and $N(z)$ are the probability density function (PDF) and cumulative distribution function (CDF), respectively, of the standard normal distribution. 

This volatility approximation, Eq.~\eqref{eq:hagan}, is the leading and first-order terms of the asymptotic expansion around a small $\vovn$. Therefore, the accuracy of the approximation noticeably deteriorates when $\vovn$ is not small. Despite various attempts to improve approximation~\citep{obloj2007fine,paulot2015asym,lorig2017lsv}, the intrinsic limitation arising from asymptotic approximation is difficult to overcome. Inaccurate option prices can also lead to inaccurate risk management. The option delta $D(K)$ implied from the volatility approximation,
\begin{equation} \label{eq:cdf_hagan}
D(K) = -\frac{\partial C(K)}{\partial K} \approx \frac{C(K-h) - C(K+h)}{2h} \quad\text{for small $h$},
\end{equation}
can deviate from the true delta significantly. 

\subsection{Existing option price representation} \label{ss:int}
\noindent
Based on the heat kernel on hyperbolic geometry (i.e., Poincar\'e half-plane), \citet{henry2005general,henry2008analysis} derives a two-dimensional integral representation of the option price under the normal SABR model, and \citet{korn2013exact} correct mistakes in the formula. Their formula, with the simplification of \citet[\S~3.5.2]{antonov2019modern}, is as follows:
\begin{equation} \label{eq:korntang}
C(K) = [F_0-K]^+ + \frac{1}{2\pi\sqrt{2}}\frac{\sigma_0}{\rhoc\vov} \int_{s_0}^\infty Q(s) ds,
\end{equation}
where $\rhoc$ is given in Eq.~\eqref{eq:rhoc}, and
$$ s_0 = \cosh^{-1} \left( \frac{\sqrt{k^2 + \rhoc^2} - \rho k}{\rhoc^2}\right)
\qtext{with} k = z + \rho = \frac{\vov}{\sigma_0}(K-F_0) + \rho.
$$
The integrand $Q(s)$ is given by
$$
Q(s) = \left[ e^{-\frac{s}{2}} N\left(-\frac{s}{2\vovn} + \vovn\right) + e^{\frac{s}{2}}N\left(-\frac{s}{2\vovn} - \vovn\right) \right]\int_{\alpha_-(s)}^{\alpha_+(s)} \frac{d\alpha}{\sqrt{\cosh s - \cosh d(\alpha)}} 
$$
where $d(\alpha)$ and $\alpha_\pm(s)$ are
\begin{gather*}
d(\alpha) = \cosh^{-1} \left( \frac12\left(\alpha + \frac1\alpha\right) + \frac{(k-\rho\alpha)^2}{2\rhoc^2 \alpha} \right),  \\
\alpha_\pm(s) = \rho k + \rhoc^2 \cosh s \pm \sqrt{\sinh^2 s - (k - \rho \cosh s)^2}.
\end{gather*}

\citet[\S~3.5.1]{antonov2015mixing} also derive a different two-dimensional integral representation:
\begin{equation} \label{eq:antonov}
C(K) = [F_0-K]^+ + \frac{\sigma_0}{\pi\vov} \int_{s_0}^\infty \frac{G(4\vovnsq, s)}{\sinh s} \sqrt{\sinh^2 s - (k - \rho \cosh s)^2}\; ds
\end{equation}
where $G(t,s)$ is the CDF of the heat kernel~\citep[\S~3.4.5]{antonov2019modern}:
$$ G(t,s) = \frac{2e^{-t/8}}{t\sqrt{\pi t}} \int_s^\infty du\, u e^{-u^2/2t} \sqrt{\cosh u - \cosh s}.
$$
This representation is also based on the hyperbolic geometry heat kernel; however, the outcome differs from that of \citet{korn2013exact} because the integration is performed in a different order among the variables.

To evaluate these integral representations, Eqs.~\eqref{eq:korntang} and \eqref{eq:antonov}, one must resort to numerical integration. However, the use of generic numerical integrals over two-dimensional indefinite regions can be an extremely slow process because it requires evaluations of dense points. \citet{korn2013exact} present the option prices evaluated for two test cases, but they do not provide the implementation details of the numerical integration.

\section{Gaussian quadrature integration scheme} \label{s:quad}
\noindent
Using the results of \citet{choi2019nsvh}, we present a new integral representation of the option price, which differs from those in Section~\ref{ss:int}. This new representation is evidently easier to evaluate using Gaussian quadratures. First, we review \citet{choi2019nsvh}'s price transition law.

\subsection{NSVh model and the closed-form price transition formula} \label{ss:nsvh}
\noindent
\citet{choi2019nsvh} introduce the NSVh model as
\begin{equation}
dF_t = \sigma_t \left(\rho\,d\Zdrift[\drift\vov/2]_t + \rhoc\,dX_t\right) \quad \text{and}\quad
\frac{d\sigma_t}{\sigma_t} = \vov\; d\Zdrift[\drift\vov/2]_t,
\end{equation}
where $Z_t$ and $X_t$ are two independent BMs, and $\Zdrift[\mu]_t = Z_t + \mu\, t$ is the BM with a drift $\mu$. By introducing the drift term of $Z_t$, the NSVh model generalizes the normal SABR; the normal SABR model is the NSVh model with $\lambda=0$. Based on Bougerol's identity in hyperbolic geometry~\citep{alili1997full}, the NSVh model admits the following closed-form transition law for $\sigma_T$ and $F_T$~\citep[Corollary~1]{choi2019nsvh}:
\begin{equation} \label{eq:nsvh_price}
 \begin{aligned}
\sigma_T &= \sigma_0 \exp\left(\vov \bar{Z}_T\right),\\
F_T &\distequal F_0 + \frac{\sigma_0\rho}{\vov}\,\left( e^{\vov \bar{Z}_T} -  e^{\drift \vov^2 T/2} \right) + \frac{\sigma_0\rhoc}{\vov} \cos\theta\, \phi\hspace{-0.25em}\left(\vov \bar{Z}_T, \vov\sqrt{R_T^2 + \bar{Z}_T^2}\right),\\
\end{aligned} \end{equation}
where
$$\bar{Z}_T =  \Zdrift[(\drift-1)\vov/2]_T \qtext{and} \phi(Z,D) = e^{Z/2} \sqrt{2\cosh D-2\cosh Z} \quad (Z\le D)$$
and $R_T$ is the two-dimensional squared Bessel process (i.e., $R_T=X_T^2+Y_T^2$ for two independent BMs, $X_T$ and $Y_T$), and $\theta \in [0, \pi]$ is a uniformly distributed random angle. As these random variables can be easily sampled, the transition law serves as a closed-form exact simulation scheme for the normal SABR model, which outperforms the method of \citet{cai2017sabr} for the $\beta=0$ case. See \citet{choi2019nsvh} for further details.

\subsection{New integral representation of the option price}
\noindent
The closed-form transition law in Eq.~\eqref{eq:nsvh_price} also enables an efficient pricing method for vanilla options. The random variables $Z_T$, $R_T$, and $\theta$ follow normal, exponential, and uniform distributions, respectively, whose PDFs are simple.

Let us introduce the integral variables $u$ and $v$, which represent $Z_T$ and $R_T$, respectively:
$$u = \frac{\Zdrift[-\vov/2]_T}{\sqrt{T}} = \frac{Z_T}{\sqrt{T}} - \vovn \quad \left( \ulam = \frac{\bar{Z}_T}{\sqrt{T}} = u + \lambda\vovn\right)\qtext{and} v = \frac{R_{T}^2}{T}.$$
Here, $u_\lambda$ is introduced to maintain the generality of the NSVh model. Under the normal SABR model (i.e., $\lambda=0$), however, $\ulam$ is equal to $u$. 
The probability densities around variables $u$, $v$, and $\theta$ are respectively given by
$$f_u(u) = e^{-\vovn u-\frac{\vovnsq}2} n(u),\quad f_v(v) = \frac12 e^{-\frac12 v}, \qtext{and} f_\theta(\theta) = \frac{1}{\pi}.$$
Note that the term $e^{-\vovn u-\frac{\vovnsq}2}$ is the Radon-Nikodym derivative that arises from our definition of $u$. We define $u$ as such because the term makes the integrand more suitable for numerical integration, as opposed to naively defining $u = Z_T/\sqrt{T}$. For example, $e^{-\vovn u}$ offsets $e^{Z/2}$ from $\phi(Z,D)$, mitigating the exponential growth as shown below.

The terminal asset price $F_T$ is expressed as a function of $u$, $v$, and $\theta$,
$$
F_T(u, v, \theta)= F_0 + \frac{\rho\sigma_0}{\vov}e^{2\drift \vovnsq} \left( e^{2\vovn u} - 1 \right) + \frac{\rhoc\sigma_0}{\vov} \cosh \theta\; \phi\hspace{-0.25em}\left(2\vovn \ulam, 2\vovn \sqrt{v + \ulam^2}\right)
$$
Therefore, the undiscounted price of the European option struck at $K$ under the normal SABR model is expressed as an expectation over the three variables:
$$
C(K) = \int_{-\infty}^\infty du\, e^{-\vovn u - \frac{\vovnsq}2} n(u) \int_0^\infty \frac{dv}{2} e^{-\frac{v}2} \int_0^\pi \frac{d\theta}{\pi} [F_T(u,v,\theta) - K]^+.
$$
We further decompose the payout by introducing
\begin{align}
k(u) &= \frac{\vov}{\sigma_0} e^{-\vovn \ulam} (K-\fwd) - 2\rho e^{\lambda \vovnsq}\; \sinh \left(\vovn u\right), \label{eq:k} \\ 
h(u,v) &= \rhoc \left( 2\cosh\left(2\vovn\sqrt{v + \ulam^2}\right) - 2\cosh(2\vovn \ulam) \right)^{1/2}. \label{eq:h}
\end{align}
We can express the payout and option price as
\begin{gather}
F_T - K = \frac{\sigma_0}{\vov}e^{\vovn \ulam}\left( h(u,v)\cos\theta - k(u) \right),\\
C(K) = \frac{\sigma_0}{\vov}e^{(\drift-1/2)\vovnsq} \int_{-\infty}^\infty du\, n(u) \int_0^\infty \frac{dv}{2} e^{-\frac{v}2} \int_0^\pi \frac{d\theta}{\pi} [h(u,v)\cos\theta - k(u)]^+.
\end{gather}

Next, we identify the integration region of positive payout. Note that, because $h(u,v)$ is a monotonically increasing function of $v$, taking values from $0$ to $\infty$ for all values of $u$, it is always possible to find $v^*(u)$ such that $h(u,v^*) = |k(u)|$:
\begin{equation} \label{eq:vstar}
v^*(u) = \frac1{4\vovnsq}\,\text{acosh}^2 \left( \cosh\left(2\vovn\ulam\right) + \frac{k^2(u)}{2\rhoc^2}  \right) - \ulam^2.
\end{equation}
Only when $k(u)=0$, $v^*(u)$ = 0.
As long as $v\ge v^*$, it is also possible to find $\theta^*$ such that $h(u,v)\cos\theta^* = |k(u)|$:
\begin{equation} \label{eq:thetastar}
\theta^*(u,v) = \text{arccos}\left(\frac{|k(u)|}{h(u,v)}\right)\quad \left(0\le \theta^* \le \frac{\pi}{2}\right).
\end{equation}
When $k(u)=0$ or $v\to\infty$, $\theta^*(u,v) = \pi/2$.

Using $v^*(u)$ and $\theta^*(u,v)$, we express the region of $(v,\theta)$, where the payout, $h(u,v)\cos\theta - k(u)$, is positive or negative, depending on the sign of $k(u)$. 
When $k(u)\ge 0$, the payout is positive, if 
$$ \{(v,\theta): 0\le v \le v^*(u) \qtext{and} 0\le \theta \le \theta^*(u,v) \}.
$$
Therefore,
\begin{align*}
&\int_0^\infty \frac{dv}{2} e^{-\frac{v}2} \int_0^\pi \frac{d\theta}{\pi} [h(u,v)\cos\theta - k(u)]^+ \\
&= \int_{v^*}^\infty \frac{dv}{2\pi}\, e^{-\frac{v}2} \int_0^{\theta^*} d\theta \, (h(u,v)\cos\theta - k(u))\\ 
&= \int_{v^*}^\infty \frac{dv}{2\pi}\, e^{-\frac{v}2} \left(\sqrt{h^2(u,v) - k^2(u)} - \theta^*(u,v)k(u) \right)
\end{align*}

When $k(u)<0$, it is easier to identify the region of negative payout:
$$ \{(v,\theta): 0\le v \le v^*(u) \qtext{and} \pi - \theta^*(u,v)\le \theta \le \pi \}.
$$
Using the identity $[x]^+ = x - [x]^-$, where $x=h(u,v)\cos\theta - k(u)$,
the integration along $\theta$ and $v$ becomes
\begin{align*}
\int_0^\infty & \frac{dv}{2} e^{-\frac{v}2} \int_0^\pi \frac{d\theta}{\pi} [h(u,v)\cos\theta - k(u)]^+ \\
&= -k(u) - \int_0^\infty \frac{dv}{2} e^{-\frac{v}2} \int_0^\pi \frac{d\theta}{\pi} [h(u,v)\cos\theta - k(u)]^- \\
&= -k(u) - \int_{v^*}^\infty \frac{dv}{2\pi}\, e^{-\frac{v}2} \int^{\pi}_{\pi-\theta^*} d\theta \, (h(u,v)\cos\theta - k(u))\\ 
&= -k(u) + \int_{v^*}^\infty \frac{dv}{2\pi}\, e^{-\frac{v}2} \left(\sqrt{h^2(u,v) - k^2(u)} + \theta^*(u,v)k(u) \right)
\end{align*}
Combining the two cases, we finally represent the option price as
\begin{equation} \label{eq:call}
\begin{aligned}
C(K)
=& \frac{\sigma_0}{\vov}e^{(\drift-1/2)\vovnsq} \int_{-\infty}^\infty du\, n(u)\left( [-k(u)]^+ + \int_{v^*}^\infty \frac{dv}{2\pi} e^{-\frac{v}2} \left(\sqrt{h^2(u,v) - k^2(u)} - \theta^*(u,v)|k(u)| \right)\right) \\
=& \frac{\sigma_0}{\vov}e^{(\drift-1/2)\vovnsq} \int_{-\infty}^\infty du\, n(u)\\
&\quad \left( [-k(u)]^+ + e^{-\frac{v^*}2} \int_{0}^\infty \frac{dv}{2\pi} e^{-\frac{v}2} \left(\sqrt{h^2(u,v^*+v) - k^2(u)} - \theta^*(u,v^*+v)|k(u)| \right)\right),
\end{aligned}
\end{equation}
where $k(u)$, $h(u,v)$, $v^*(u)$, and $\theta^*(u,v)$ are defined in 
Eqs.~\eqref{eq:k}, \eqref{eq:h}, \eqref{eq:vstar}, and \eqref{eq:thetastar}, respectively.

The option delta, $D(K)=\text{Prob}(F_T\ge K)$, can be similarly obtained as
\begin{equation} \label{eq:delta}
\begin{aligned}
D(K) &= \int_{-\infty}^\infty du\, e^{-\vovn u - \frac{\vovnsq}2} n(u) \int_0^\infty \frac{dv}{2\pi} e^{-\frac{v}2} \int_0^\pi d\theta\, \boldsymbol{1}_{h(u,v)\cos\theta > k(u)} \\
&= \int_{-\infty}^\infty du\, e^{-\vovn u - \frac{\vovnsq}2} n(u) \left( \boldsymbol{1}_{-k(u)} + \sgn(k(u))\int_{v^*}^\infty \frac{dv}{2\pi} e^{-\frac{v}2}\,\theta^*(u,v)\right)\\
&= \int_{-\infty}^\infty du\, e^{-\vovn u - \frac{\vovnsq}2} n(u) \left(\boldsymbol{1}_{-k(u)} + \sgn(k(u))e^{-\frac{v^*}2} \int_0^\infty \frac{dv}{2\pi} e^{-\frac{v}2}\,\theta^*(u,v^* + v) \right),
\end{aligned}
\end{equation}
where $\sgn(x)$ is the sign function and $\boldsymbol{1}_x$ is the indicator function (i.e., $\boldsymbol{1}_x = 1$ if $x>0$ or 0 otherwise). From option delta $D(K)$, the CDF at $K$ can also be obtained as $ F(K) = 1 - D(K)$.

\subsection{Integration with Gaussian quadratures}
\noindent
The integral representations in the previous section allow for efficient evaluations because Gaussian quadratures exist for the probability densities of the variables. While we integrate $\theta$ analytically, we use the Gauss-Hermite quadrature for $u$ and the Gauss-Laguerre quadrature for $v$. Let $\{u_i\}$ and $\{w_i\}$ for $i=1,\ldots,N$ be the points and weights, respectively, of the Gauss-Hermite quadrature with respect to the weight function $n(u)$, and let $\{v_j\}$ and $\{\bar{w}_j\}$ for $j=1,\ldots,M$ be those of the Gauss--Laguerre quadrature associated with the weight function $e^{-\frac{v}2}$ ($v\ge 0$). The integrations with respect to $n(u)$ and $e^{-v/2}$ are then accurately approximated by the weighted sums,
\begin{equation*}
\int_{-\infty}^\infty du\; n(u) f(u) \approx \sum_{i=1}^N w_i f(u_i) \qtext{and}\int_{v^*}^\infty dv\; e^{-\frac{v}{2}} g(v) \approx e^{-\frac{v^*}{2}} \sum_{k=1}^M \bar{w}_j g(v^* + v_j),
\end{equation*}
for some functions $f(u)$ and $g(v)$, respectively. The points and weights of the Gaussian quadrature are optimally chosen from orthogonal polynomials, in the sense that the quadrature of size $N$ can precisely evaluate the moments of the probability density up to the order $2N-1$. The quadrature points and weights can be easily generated using public numerical libraries.\footnote{We use the python functions,  \href{https://docs.scipy.org/doc/scipy/reference/generated/scipy.special.eval_hermitenorm.html}{\texttt{scipy.special.eval\_hermitenorm}} and \href{https://docs.scipy.org/doc/scipy/reference/generated/scipy.special.roots_genlaguerre.html}{\texttt{scipy.special.roots\_genlaguerre}}, for Gauss--Hermite and Gauss--Laguerre quadratures, respectively.} Because we perform two-dimensional integration, we construct the compound quadrature $(u_i, v_j)$ with weight $w_i\bar{w}_j$, whose size is $NM$.

Let us define the indexed function values at the quadrature points by
$$
k_i = k(u_i), \quad v^*_i = v^*(u_i), \quad h_{ij} = h(u_i, v^*_i + v_j),\quad \theta^*_{ij}=\theta^*(u_i,v^*_i + v_j).
$$
Then, the option price in Eq.~\eqref{eq:call} is evaluated as a double sum over the composite Gaussian quadratures:
\begin{equation} \label{eq:quad_opt}
C(K) = \frac{\sigma_0}{\vov} e^{(\drift-1/2)\vovnsq} \sum_{i=1}^N  w_i \left([-k_i]^+ + e^{-\frac{v^*_i}{2}} \sum_{j=1}^M \frac{\bar{w}_j}{2\pi} \left(\sqrt{h_{ij}^2-k_i^2} - \theta^*_{ij}\,|k_i|\right) \right).
\end{equation}
The calculation can be performed simply as a vector--matrix operation:
$$
C(K) 
= \frac{\sigma_0}{\vov} e^{(\drift-1/2)\vovnsq} 
\boldsymbol{w}^T \left(\boldsymbol{b} + \boldsymbol{A} \boldsymbol{\bar{w}}\right),
$$
where $\boldsymbol{w}$ is the size $N$ vector of $w_i$, $\boldsymbol{\bar{w}}$ is the size $M$ vector of $\bar{w}_j$, $\boldsymbol{b}$ is the size $N$ vector of $[-k_i]^+$, and $\boldsymbol{A}$ is the $N\times M$ matrix of 
$$ A_{ij} = \frac{1}{2\pi} e^{-\frac{v^*_i}{2}} \left[\sqrt{h_{ij}^2-k_i^2} - \theta^*_{ij}\,k_i\right].
$$
The option delta in Eq.~\eqref{eq:delta} can be similarly evaluated using the Gaussian quadrature:
$$
D(K) = \sum_{i=1}^N \left[w_i e^{-\vovn u_i - \frac{\vovnsq}2}  \left(\boldsymbol{1}_{-k_i} + \sgn(k_i) e^{-\frac{v^*_i}{2}}\sum_{j=1}^M \frac{\bar{w}_j}{2\pi} \theta^*_{ij}\right) \right].
$$

\section{Numerical Experiments} \label{s:num}
\begin{table} 
\caption{Parameter sets used in the simulations} \label{t:param}
\begin{center}
	\begin{tabular}{|c||c|c|c|c|c|} \hline 
	Case & $\sigma_0$ & $\vov$ & $\rho$ & $T$ & $F_0$ \\ \hline\hline
1 & 100 & 0.5 & 0 & 30 & 350 \\
2 & 100 & 0.5 & -0.3 & 30 & 350 \\
3 & 100 & 0.5 & -0.6 & 30 & 350 \\
	\hline
	\end{tabular}
\end{center}
\end{table}

\noindent
We test the accuracy of our Gaussian quadrature method for calculating the option price and delta. We use the three parameter sets shown in Table~\ref{t:param}. Case 1 is one of the parameter sets tested by \citet[Tables 3--4 and Figure 4]{korn2013exact}, but scaled by $10^4$. This case is understood as a parameter set for the swaption volatility smile in the unit of basis point. We price this case again because it is a challenging example. Given a large value of $\vovn=1.37$, the error from Hagan's normal volatility approximation in Eq.~\eqref{eq:hagan} is significantly large, as shown by \citet{korn2013exact}. From Case 1, we vary correlation $\rho$ from 0\% to $-30\%$ in Case 2 and $-60\%$ in Case 3.

\begin{table}[ht]
\caption{\label{t:case1} The call option price and delta for Case 1 obtained from $N\times M$ Gaussian quadrature points and Hagan's volatility approximation. The exact values are obtained with $90\times 180$ points. The delta value and error are in \% unit.}
\begin{center} \small
\begin{tabular}{c||c|c|c|c|c||c|c|c|c|c} \hline
	Case 1 & \multicolumn{5}{c||}{Option price} & \multicolumn{5}{c}{Option delta (\%)} \\ \hline
	& \multicolumn{4}{c|}{Error from the exact value} & Exact & \multicolumn{4}{c|}{Error from the exact value} & Exact \\ \cline{2-5}\cline{7-10}
	Strike & 7$\times$7 & 10$\times$10 & 14$\times$14 & Hagan & value & 7$\times$7 & 10$\times$10 & 14$\times$14 & Hagan & value \\ \hline
0 & ~0.16 & 0.14 & 0.07 & 92.28 & 572.02 & -0.19 & -0.10 & -0.06 & ~21.45 & 84.47 \\
100 & ~0.21 & 0.15 & 0.08 & 70.49 & 489.88 & -0.26 & -0.14 & -0.09 & ~21.85 & 79.42 \\
200 & ~0.30 & 0.17 & 0.10 & 49.61 & 414.24 & -0.41 & -0.23 & -0.14 & ~19.03 & 71.16 \\
300 & ~0.45 & 0.27 & 0.16 & 35.07 & 349.19 & -0.64 & -0.44 & -0.31 & ~8.43 & 58.06 \\
350 & -0.01 & 0.00 & 0.00 & 32.92 & 322.16 & -0.00 & -0.00 & ~0.00 & -0.00 & 50.00 \\
400 & ~0.45 & 0.27 & 0.16 & 35.07 & 299.19 & ~0.64 & ~0.44 & ~0.31 & -8.43 & 41.94 \\
500 & ~0.30 & 0.17 & 0.10 & 49.61 & 264.24 & ~0.41 & ~0.23 & ~0.14 & -19.03 & 28.84 \\
600 & ~0.21 & 0.15 & 0.08 & 70.49 & 239.88 & ~0.26 & ~0.14 & ~0.09 & -21.85 & 20.58 \\
700 & ~0.16 & 0.14 & 0.07 & 92.28 & 222.02 & ~0.19 & ~0.10 & ~0.06 & -21.45 & 15.53 \\
	\hline
\end{tabular} \vspace{1em}
\end{center}
\end{table}

\begin{table}[ht]
\caption{\label{t:case2} The call option price and delta for Case 2 obtained from $N\times M$ Gaussian quadrature points and Hagan's volatility approximation. The exact values are obtained with $90\times 180$ points. The delta value and error are in \% unit.}
\begin{center} \small
\begin{tabular}{c||c|c|c|c|c||c|c|c|c|c} \hline
	Case 2 & \multicolumn{5}{c||}{Option price} & \multicolumn{5}{c}{Option delta (\%)} \\ \hline
	& \multicolumn{4}{c|}{Error from the exact value} & Exact & \multicolumn{4}{c|}{Error from the exact value} & Exact \\ \cline{2-5}\cline{7-10}
	Strike & 7$\times$7 & 10$\times$10 & 14$\times$14 & Hagan & value & 7$\times$7 & 10$\times$10 & 14$\times$14 & Hagan & value \\ \hline
0 & ~0.22 & 0.21 & 0.10 & 105.57 & 580.55 & -0.13 & -0.09 & -0.06 & ~22.48 & 86.50 \\
100 & ~0.28 & 0.19 & 0.11 & 82.14 & 495.84 & -0.20 & -0.09 & -0.07 & ~24.33 & 82.66 \\
200 & ~0.39 & 0.14 & 0.08 & 57.25 & 415.99 & -0.30 & -0.12 & -0.08 & ~25.09 & 76.51 \\
300 & ~0.48 & 0.27 & 0.14 & 33.37 & 344.19 & -0.45 & -0.24 & -0.15 & ~21.48 & 66.23 \\
350 & ~0.46 & 0.36 & 0.19 & 23.83 & 312.82 & -0.53 & -0.36 & -0.24 & ~16.24 & 59.01 \\
400 & ~0.52 & 0.29 & 0.18 & 17.55 & 285.36 & -0.13 & -0.10 & -0.08 & ~8.54 & 50.68 \\
500 & ~0.53 & 0.26 & 0.12 & 17.29 & 243.03 & ~0.56 & ~0.35 & ~0.22 & -7.05 & 34.52 \\
600 & ~0.34 & 0.16 & 0.07 & 28.63 & 214.53 & ~0.31 & ~0.17 & ~0.10 & -14.24 & 23.41 \\
700 & ~0.25 & 0.13 & 0.05 & 43.88 & 194.70 & ~0.20 & ~0.11 & ~0.06 & -15.70 & 16.83 \\
	\hline
\end{tabular} \vspace{1em}
\end{center}
\end{table}

\begin{table}[ht]
\caption{\label{t:case3} The call option price and delta for Case 3 obtained from $N\times M$ Gaussian quadrature points and Hagan's volatility approximation. The exact values are obtained with $90\times 180$ points. The delta value and error are in \% unit.}
\begin{center} \small
\begin{tabular}{c||c|c|c|c|c||c|c|c|c|c} \hline
	Case 3 & \multicolumn{5}{c||}{Option price} & \multicolumn{5}{c}{Option delta (\%)} \\ \hline
	& \multicolumn{4}{c|}{Error from the exact value} & Exact & \multicolumn{4}{c|}{Error from the exact value} & Exact \\ \cline{2-5}\cline{7-10}
	Strike & 7$\times$7 & 10$\times$10 & 14$\times$14 & Hagan & value & 7$\times$7 & 10$\times$10 & 14$\times$14 & Hagan & value \\ \hline
0 & -0.04 & 0.04 & 0.03 & 72.67 & 569.45 & -0.03 & -0.09 & -0.07 & ~16.99 & 89.20 \\
100 & -0.04 & 0.13 & 0.01 & 54.56 & 481.52 & -0.13 & -0.04 & -0.04 & ~19.30 & 86.47 \\
200 & ~0.04 & 0.14 & 0.06 & 33.96 & 397.03 & -0.25 & -0.03 & -0.02 & ~21.92 & 82.16 \\
300 & ~0.13 & 0.13 & 0.06 & 10.93 & 318.23 & -0.33 & -0.09 & -0.06 & ~23.80 & 74.72 \\
350 & ~0.05 & 0.14 & 0.07 & -0.91 & 282.24 & -0.32 & -0.18 & -0.11 & ~23.26 & 68.93 \\
400 & ~0.30 & 0.18 & 0.09 & -11.93 & 249.61 & -0.39 & -0.30 & -0.19 & ~20.32 & 61.23 \\
500 & ~0.40 & 0.22 & 0.11 & -25.71 & 198.02 & ~0.67 & ~0.48 & ~0.33 & ~5.97 & 41.57 \\
600 & ~0.14 & 0.05 & 0.02 & -25.13 & 165.13 & ~0.33 & ~0.19 & ~0.11 & -5.36 & 25.56 \\
700 & ~0.07 & 0.02 & 0.01 & -17.86 & 144.45 & ~0.18 & ~0.10 & ~0.06 & -8.32 & 16.74 \\
	\hline
\end{tabular} \vspace{1em}
\end{center}
\end{table}

\begin{figure}[ht!]
	\caption{\label{f:1} The volatility smile (left) and delta (right) as functions of strike price for the three test cases in Table~\ref{t:param}.}
	\includegraphics[width=0.49\textwidth]{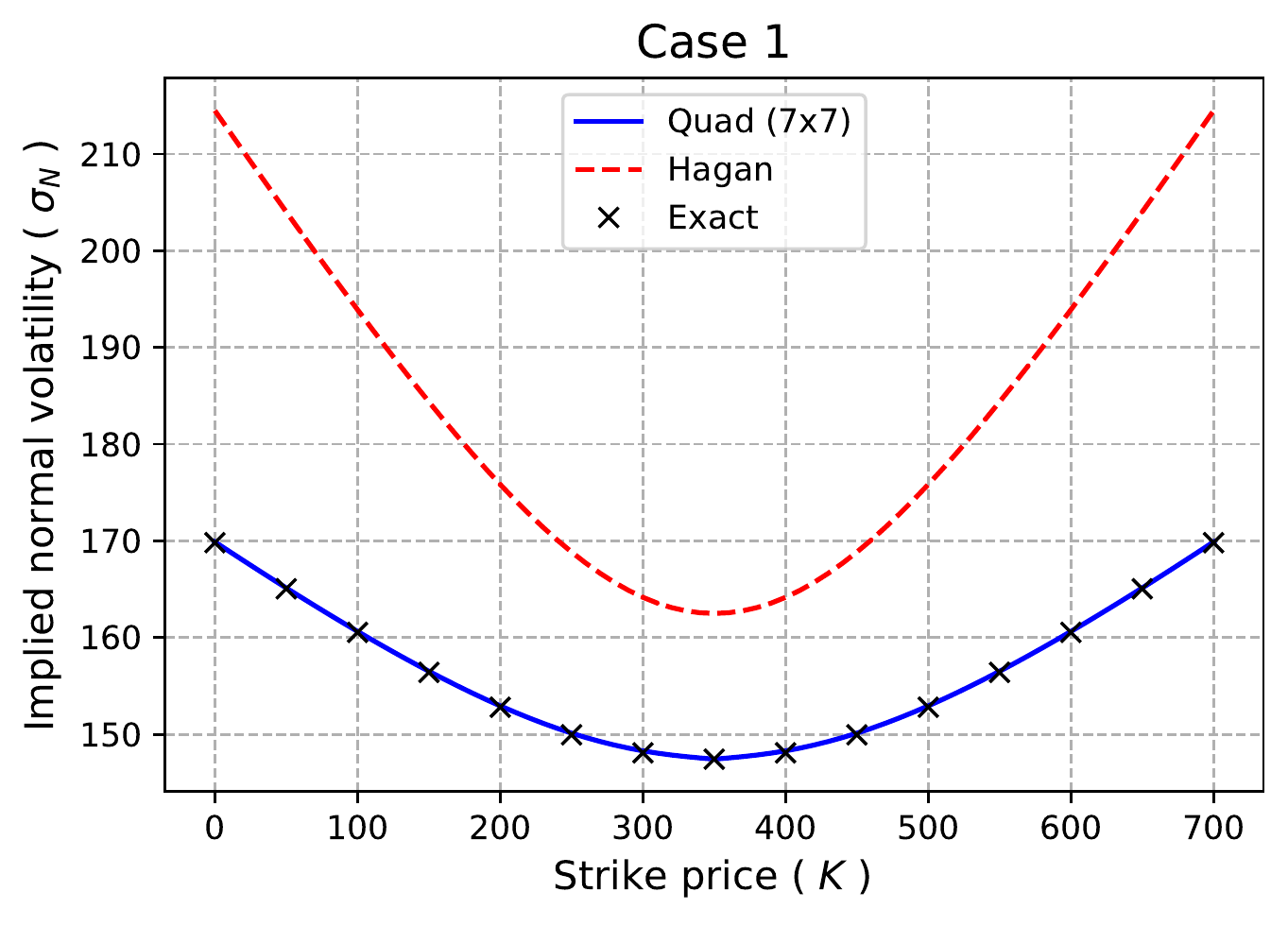}
	\includegraphics[width=0.49\textwidth]{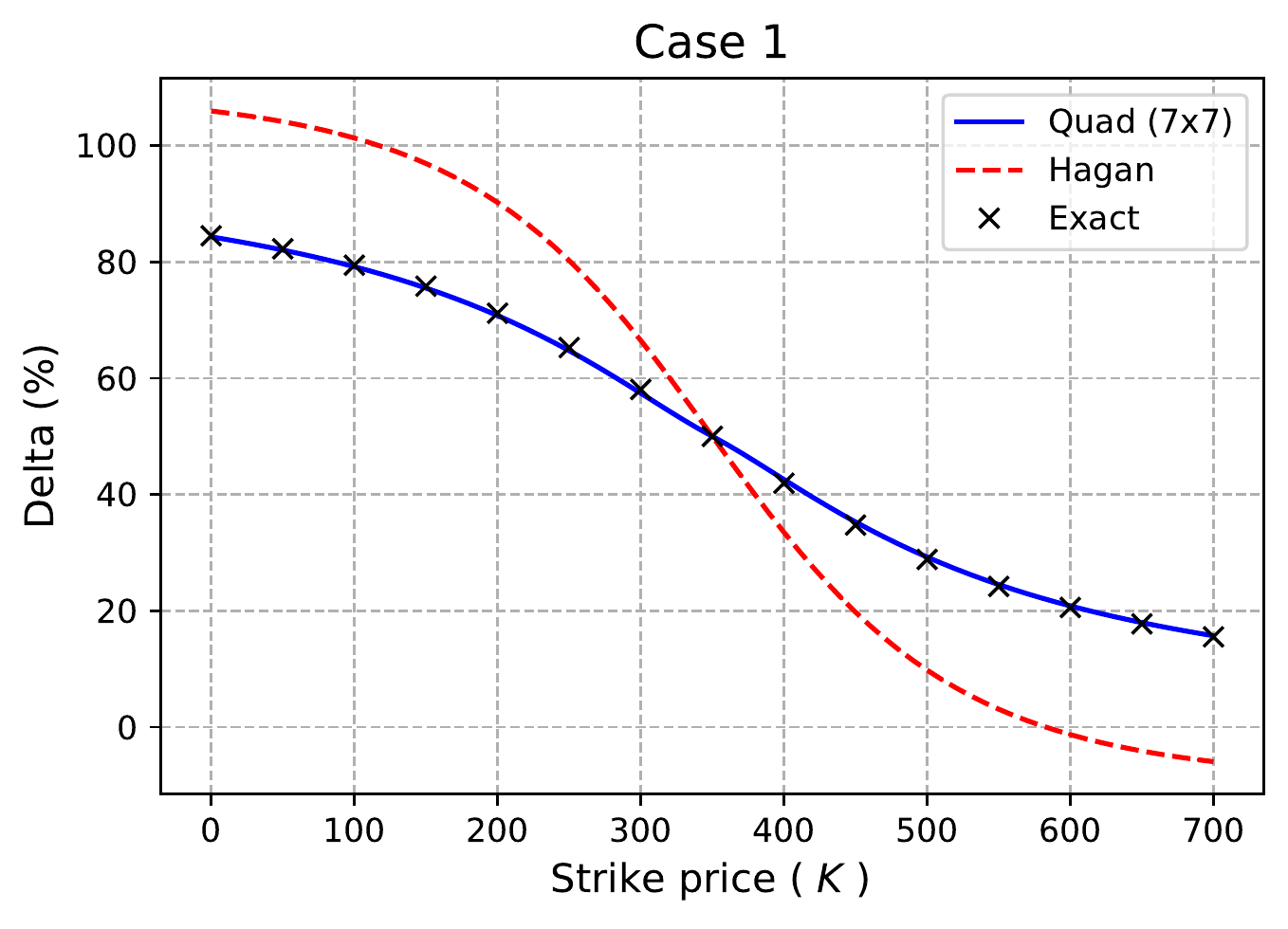}\\ \vspace{1ex}
	\includegraphics[width=0.49\textwidth]{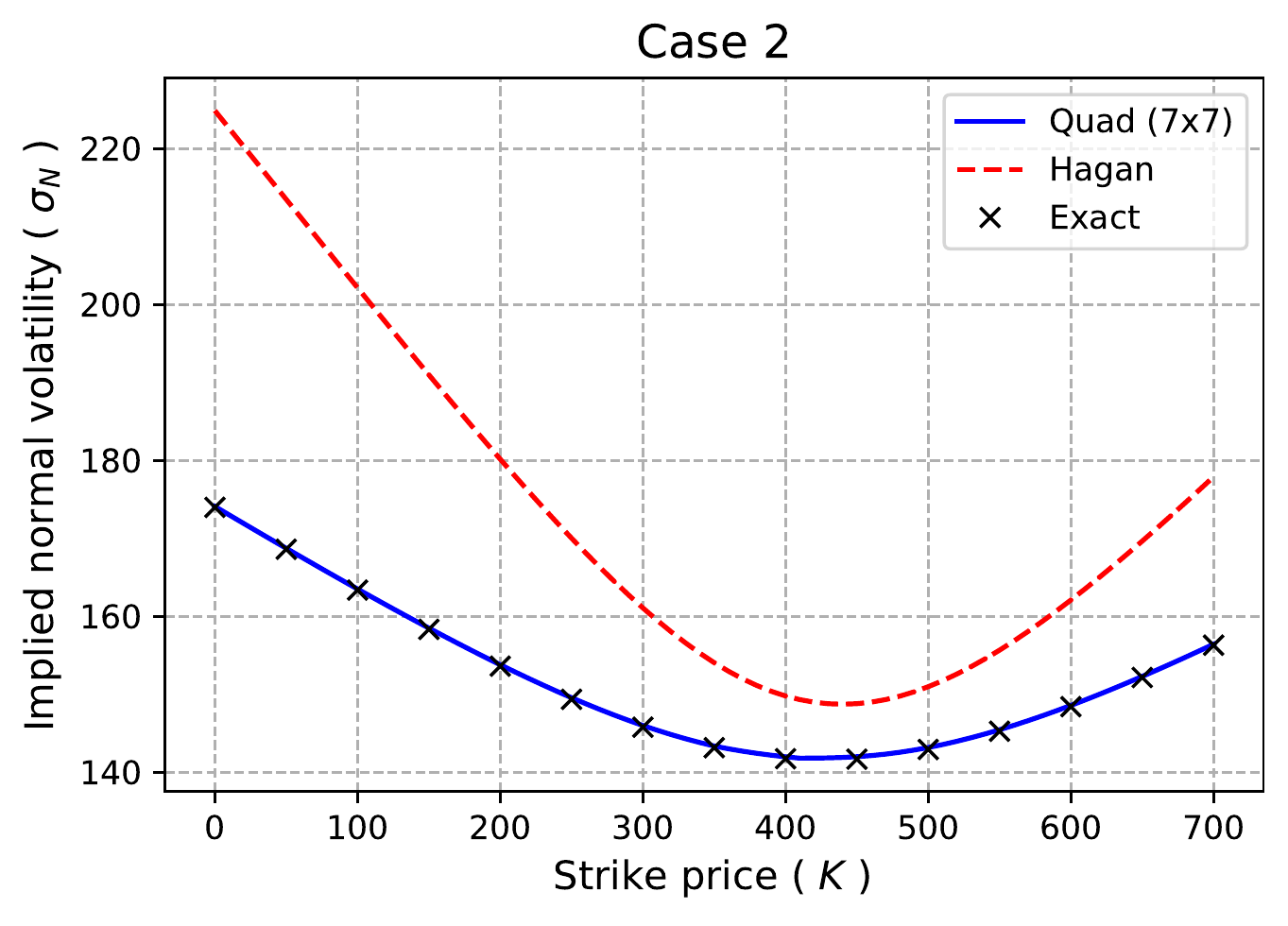}
	\includegraphics[width=0.49\textwidth]{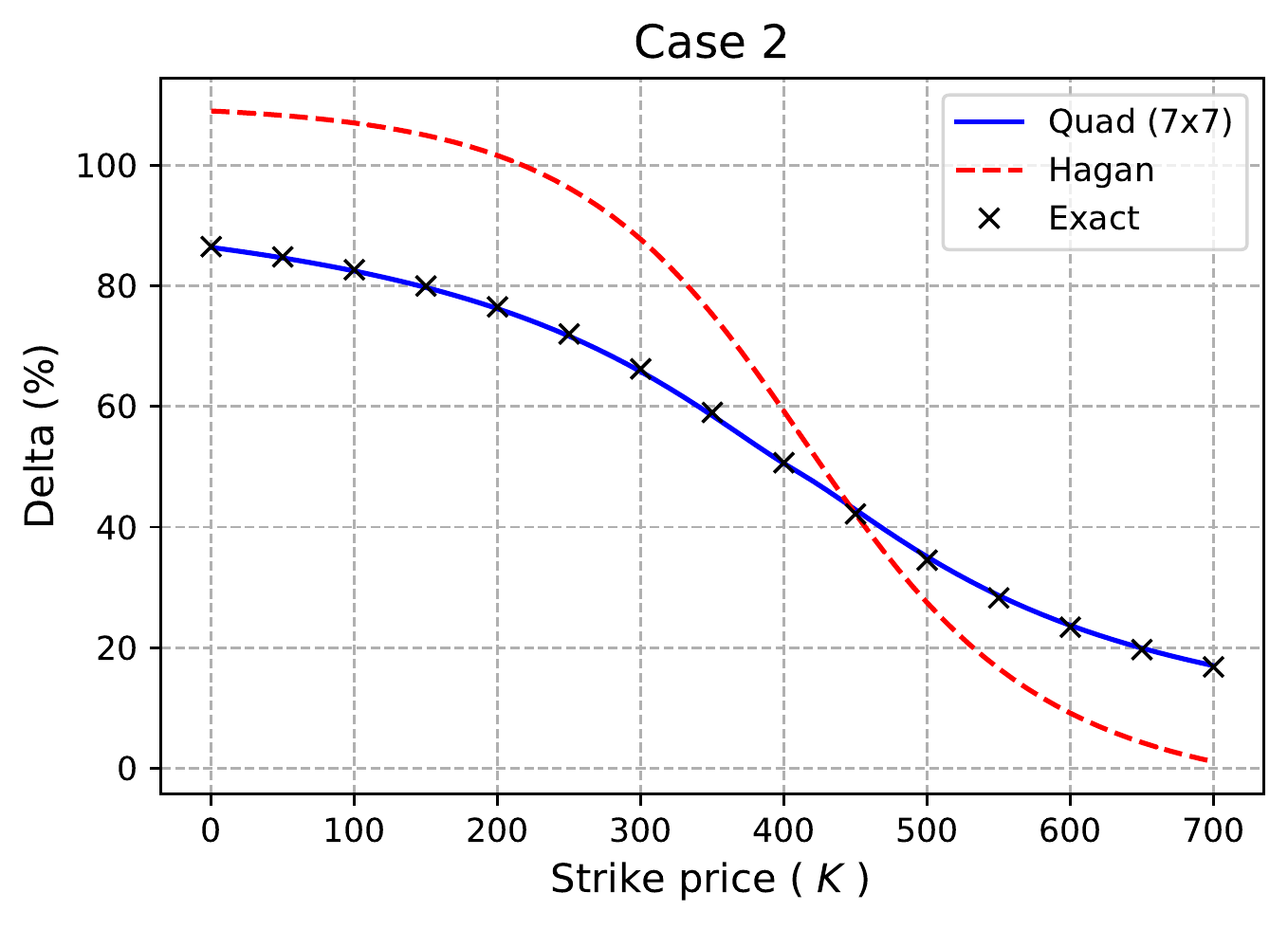}\\ \vspace{1ex}
	\includegraphics[width=0.49\textwidth]{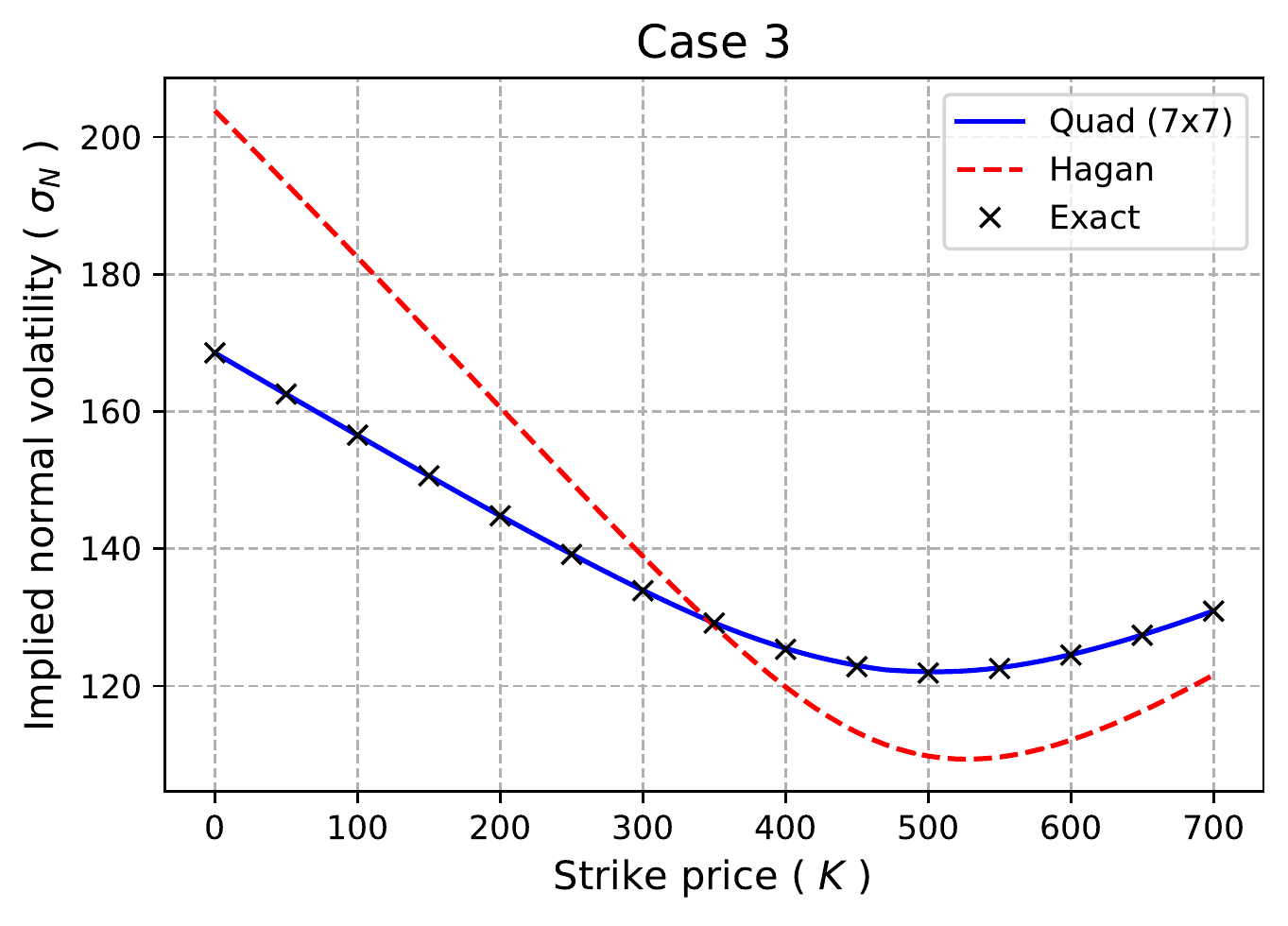}
	\includegraphics[width=0.49\textwidth]{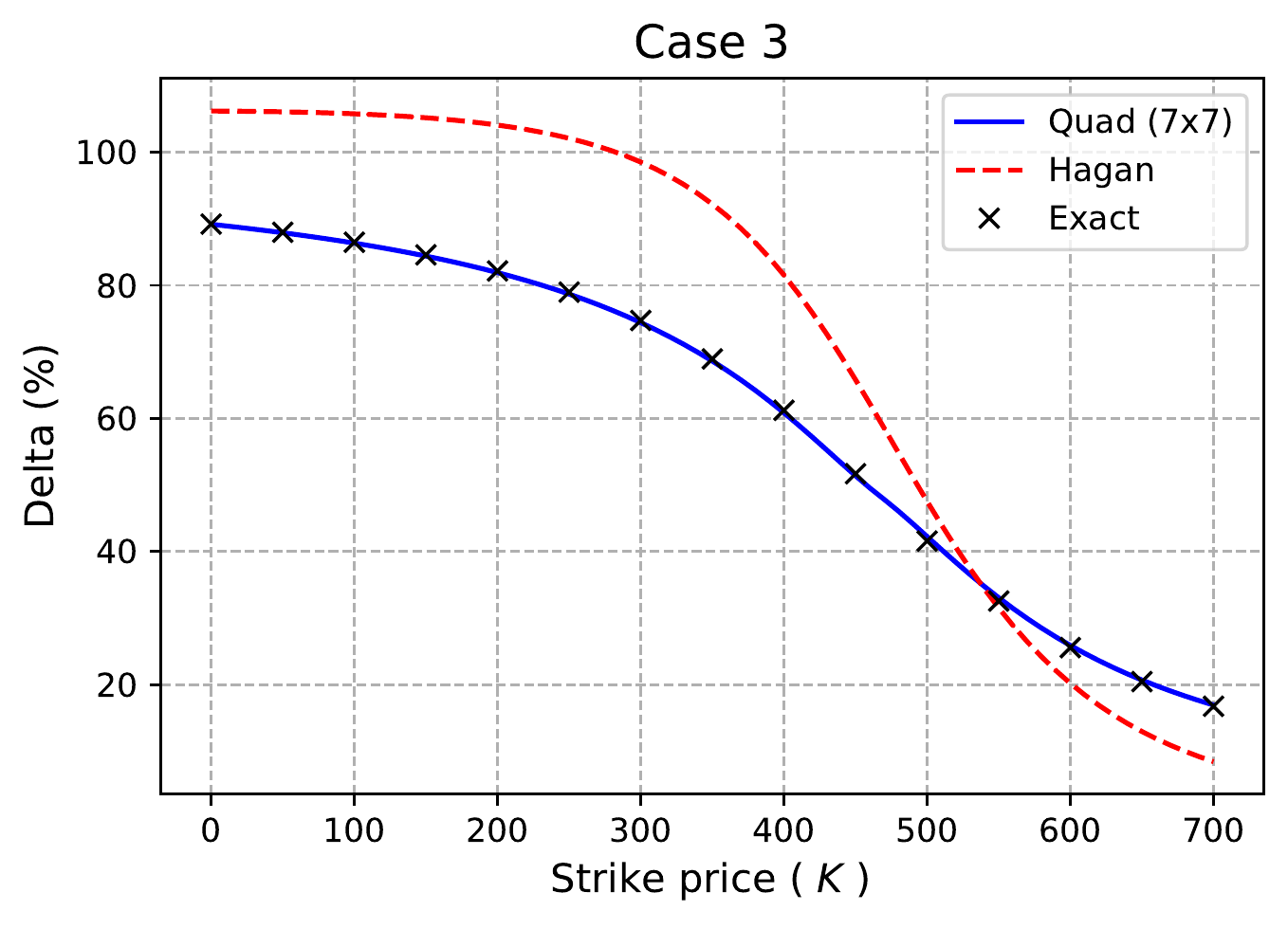}\\ \vspace{1ex}
\end{figure}

Tables~\ref{t:case1}--\ref{t:case3} present the results for the three cases. We first obtain the exact values of option price and delta from the Gaussian quadrature integration (Eqs.~\eqref{eq:call} and \eqref{eq:delta}, respectively) with a very dense quadrature set with $N=90$ and $M=180$. The option prices for $K=300$, 350, and 400 in Case 1 is consistent with the values reported by \citet[Table~3]{korn2013exact}. Then, we report the errors of the option price and delta from a small quadrature size and from Hagan's volatility approximation (Eqs.~\eqref{eq:p_hagan} and \eqref{eq:cdf_hagan}). We increase the quadrature size $N=M$ from  7, to 10, to 14, roughly doubling the total number of points from 49, to 100, to 196. The tables show that our quadrature integration accurately evaluates the price and delta. With only $7\times 7$ points, the price error is less than one basis point and the delta error is within 1\%. These errors are negligible for practical purposes. Figure~\ref{f:1} displays the implied normal volatility (left) and delta (right) for the three test cases. The result with $7\times 7$ points is visually indistinguishable from the exact value with $90\times 180$ points.
By contrast, Hagan's volatility approximation shows significant deviations from the exact value. Moreover, the option delta incorrectly goes below 0 or above 100\%. Given that the option delta is equal to the complementary CDF under the normal SABR model, this implies an arbitrage opportunity. It illustrates a possible danger of using a volatility approximation approach.

\section{Conclusion} \label{s:conc}
\noindent
The normal ($\beta=0$) SABR model is an important special case of the popular SABR model for modeling interest rates as it allows negative asset prices. Although Hagan's implied volatility approximation is easy to use, its accuracy deteriorates and arbitrage is possible. This study provides an efficient numerical scheme for option prices and deltas. Based on the new price transition law, our proposed scheme adopts a compound Gaussian quadrature. Numerical tests show that a quadrature with only $7\times 7$ nodes accurately evaluates European options without arbitrage. Given that the SABR model still requires better pricing methods for general cases ($0\le \beta\le 1$), our new scheme will serve to provide the benchmarks for testing the methods proposed in the future.

\section*{Declarations of Interest}
\noindent The authors report no conflicts of interest. The authors alone are responsible for the content and writing of the paper.

\begin{singlespace}
\bibliography{../../@Bib/SABR}
\end{singlespace}
\end{document}